# Huge Ballistic Magnetoresistance in Multiple Nanocontacts Devices

*N. García[1], M.R. Ibarra [2,3], C. Hao[1], R. F. Pacheco[2] and D. Serrate[3]*

[1] *Laboratorio de Física de Sistemas Pequeños y Nanotecnología*
*Consejo Superior de Investigaciones Científicas, Serrano 144, Madrid 28006, Spain*
[2] *Instituto de Nanociencia de Aragón INA, Universidad de Zaragoza, Edificio Interfacultativo II, Pedro Cerbuna 12, 50009-Zaragoza, Spain*
[3] *Instituto de Ciencia de Materiales de Aragón ICMA Universidad de Zaragoza-Consejo Superior Investigaciones Científicas, Facultad de Ciencias, Pedro Cerbuna 12, 50009-Zaragoza, Spain*

**ABSTRACT**
In this paper we report an exhaustive experimental work on magnetoresistance effects found in a system in which a large number of nanocontacts are produced between oxidized Fe fine particles. We have obtained the following performances: i) Huge low field room temperature magnetoresistance (over 1000%). ii) Non-linear I-V at different applied fields and temperatures. iii) Large thermal stability and reproducible resistance value under thermal cycles from room temperature down to 5 K. iv) Easy to fabricate with an almost 100% success. v) Heavy duty and transportable samples with reproducibility tested in several laboratories. We realized that the extraordinary effect found is related to the oxygen content at the particles surface.

Magnetotransport phenomena based on the existence of large magnetoresistance (MR) at room temperature have received lately much attention in research and development due to applications in spintronic and in a not less important field of the contact less sensors. Most of the current applications are based on the thin film technology. Giant magnetoresistance (GMR) found in magnetic multilayers [1,2] had a large impact in technological applications. Tunnelling magnetoresistance using non magnetic oxide to define a tunnel barrier between two magnetic films is another field of research with expectation to extend the application of MR to spintronic devices [3]. The oxide grown has been in most of the cases alumina. These, use the idea proposed by Julliere [4] to explain the data and the non-magnetic oxide just play the role of a tunnel barrier and does not select spin, or better does not contribute to extra polarization of the current. There has been also other tentative with different oxides and electrodes to study the role of the metal-oxide interface on the spin polarization of the tunnel electrons [5]. Recently, larger values of room temperature tunnel magnetoresistance (around 200 %) have been reported using MgO insulator barriers [6,7].

Other well explored source of magnetoresistance is the spin dependent transport in inhomogeneous granular thin films ferromagnets [8,9]. Metallic ferromagnetic particles embedded in either insulator (tunnel or activation mechanisms) or metallic (disorder interfacial scattering) showed that large magnetoresistance appear close to the percolation threshold and does not depend on the resistance value [10]. This was explained within the framework of Hellman&Abeles theory [11]. Magnetoresistance experiments performed in compacted half ferromagnet oxides powders showed small values at room temperature [12], even under applied field up to 40 Tesla [13]. Large values of magnetoresistance, up to thousands percent, have been found by some of the authors in this paper with ballistic magnetoresistance (BMR) using point contacts of a few atoms (with resistances of 1000´s ohms)[14]. This was explained in terms of domain wall scattering [15,16]. However very large values have been also obtained in electrodeposited nanocontacts [17,18]. In this case the resistance is a few ohms, which indicates that the formed contacts are of the order of 10 nanometres diameter or that several contacts are formed. In such a case the magnetoresistance cannot be explained considering domain wall scattering [15,16,19] but to the existence of an interface layer at the contact (probably an oxide) that is highly spin polarized and acts as a filter for the spin of the electrons [19]. These nanocontacts are not very stable in time. Nevertheless, reproducible effects can be obtained over a week by immersion of the contact in oil [19]. The reason for this instability is ascribed to a deterioration of the contact due to extra oxidation which could change the nature of the interface layer. An additional problem is the high density of current flowing through the contact, which could modify or break the contact.

To avoid the problems found in individual contacts, we have imagined and performed experiments in which the current flow through a large number of nanocontacts. This has as a consequence a drastic reduction of the current density at the individual nanocontacts. We consider that the magnetotransport phenomena in our system may be dominated by an average statistical number of contacts. This in turn does not change the overall magnetoresistance. The nanocontacts are formed at the particle junctions; as a consequence the nature of the oxide at the surface is relevant. In addition the contacts should be completely encapsulated and isolated from the environment to avoid additional oxidation. We have succeeded and obtained such a system in a multiple nanocontacts device (MUND) by confining the iron fine particles between two metallic electrodes. The system is stable, robust, reproducible, easy to handle and fabricate, with huge values of room temperature and low field magnetoresistance (around 1000%). We consider that this device could be relevant for magnetoelectronic applications.

Starting micrometric (1-6 µm) powders from Goodfellow-ALFA were ball milled under different atmospheres (vacuum of $10^{-4}$ torr, ambient and $N_2$). Powders were mechanically milled during times ranging from minutes to 330 hours. Samples at selected intermediate milling times were obtained. Powders milled over hundred hours reached nanometre particle sizes around 40 nm as revealed the X-ray diffraction analysis. Small amounts of milled powders were inserted in a cylindrical hole as schematized in Fig.1a. The column of powder was compressed over a metallic plate by a metallic pin and then glued. The MR was measured at different temperatures, injected currents and voltages. SEM and EDAX analysis were performed on the different particles batches in order to determine the morphology and the overall oxygen content. This was determined by a Philips SEM FEG XL 30, at 20keV energy of the electrons and with a 5-10% accuracy determination

The MR obtained varies depending of the milling time and the ambient in which the milling is performed. The typical result for particles obtained after 200 hour milling time (average size 40nm) under vacuum of $10^{-4}$ Torr is over 1000% MR. Under these conditions we obtained the best results. Also other milling conditions under oxygen and nitrogen atmosphere and other milling times have been used. In Fig. 1b we can appreciate typical magnetoresistance loops and the resistance dependence with the measuring current at 0 and 6000Oe. It is remarkable, first the large values of the magnetoresistance and the reproducibility of the MR loops and also the considerable dependence with the current. Te resistance varies strongly for zero field and very little for high field, which is an indication that we are dealing with a magnetoelectronic effect.

For each badge of milled powder that we have used, in our more than 700 samples preparations, we have obtained SEM micrographs and EDAX analysis to determine the overall oxygen content. Huge MR values were found using the powders milled under vacuum. However the values obtained in samples milled at ambient conditions were negligibles. Based on these results and considering the previous results on the role of the oxygen at the electrodeposited nanocontacts [19] we considered relevant to analyze the oxygen content. Indeed we have found that large MR values are only achieved within a limited oxygen content range between 2.5 – 3.3% (0.25%) (Fig. 2). Indeed, we have observed that a large peak in MR appears for samples milled around 200 hours. The MR value within this region is reduced as the time of milling is reduced or increased. We also have done HRTEM as well as high resolution EELS on the nanoparticles after milling and found that indeed the particles are oxidized at the surface and show crystalline disorder in a nanometre shell (5-10 nm), but the particle core is well crystallized (details on morphology and chemical composition will be reported elsewhere). We can not obtain direct information on the oxides that form the contacts; they probably are non stoichiometric. What it is clear from our results is the relevant role of the oxygen content and perhaps the change in the particle morphology as well as the strong atomic disorder at the particles surface induced by the mechanical milling. We notice also that the coercive field and hysteresis obtained from the maximum of the MR measurements is very large (typically 80 to 100 Oe) respect to that obtained from magnetization isotherms (20 Oe). This indicates that the nature of the process responsible for the huge MR is not related to the overall magnetization process in the system, which is basically due to the particles core material. As other relevant feature, we observed a large sensitivity of the MR to the injected current (Fig.1b). All these data are at RT and were measured in Madrid with powders prepared in Zaragoza. But the same results were obtained with samples prepared in Madrid and sent to Zaragoza and vice versa back and forth. Experiments are reproducible and samples transportable without precautions.

We also have measured I-V curves at different temperatures. In Fig. 3 we present typical data for 103 and 275K for I-V curves and R(H) loops. Results at low temperature look similar to those found in tunnelling processes and metallic contacts with a limited number of conduction channels, but non-linear behaviour is also found at 275K. The resistance changes by orders of magnitude with temperature but the MR remains in the same order of 1000´s% (see inset of Fig 3). We believe the transport is dominated by activation, in the range of temperature above 75K, to almost spin polarized levels where the electron spin goes ballistic, i.e in a non adiabatic way conserving spin. It should be said that in certain cases we have measured effect up to 50000%.

Now we proceed by reporting our studies of reproducibility of R(H) in time. These experiments started approximately one year ago and the samples are still working and have been transferred to other laboratories and measured [20]. We have chosen one of the samples and studied extensively since 4 months ago. Fig. 4 shows MR data for a sample measured within a 4 months period. These values, ratios of resistances, are very reproducible. The data presented in these loops are for increments of field of 100 Oe but the same results are obtained for 2 Oe increments. The inset of Fig. 4 shows the time dependence of the noise to signal ratio for a period up to 60000 seconds. Measurements were performed at two selected applied field values: the coercive field and 6000 Oe. Remarkable values of 0.1 to 0.3% were found. Fig. 5 shows the thermal variation of the resistance cycling the sample from room temperature down to 10 K. The reproducibility of the results indicates the large stability of the sample.

We have prepared also samples with other Goodfellow starting powders as Co, Ni, permalloy, CoNi, CoFe and magnetite and the MR are smaller that 5-10%. Except for permalloy that in some cases after 100´s hours milling we achieved 70%, however the R(H) curves do not follow the magnetization in permalloy as happens with the Fe particles. We do not discard that large and reproducible MR results could be obtained with the above particles using other routes. The key point in our opinion is to provide *the suitable oxide or other surface material to realize adequate contacts*. Different ways of particle fabrications produce different results, this is our observation. In fact it has been observed large MR in nanocontacts of Ni, Co, and magnetite with other sample preparations [14,16,21,22].

The observed magnetoresistive effect, using the preparation method discussed above, is characterized by the following features: i)the huge MR is only found using fine Fe particles at given values of oxygen content and milling time, ii) it has been observed only in iron particles and no in others ferromagnetic particles that have practically the same magnetizations and coercive fields, iii) the MR values strongly depends of the injected current or applied voltage, iv) the results are highly reproducible, v) there exists a very low level noise to signal ratio, and vi) the thermal dependence of resistance is uniform and reproducible. All of these performances discard possible artefacts due to magnetostriction or magnetomechanical forces which will not account for the previous results. We have also measured switching times by using pulsed current and found that they are smaller than $10^{-4}$ second that is our resolution.

In conclusion, we are measuring spintronic effects. Our results can be explained considering that electrons are excited to fully spin polarized levels. This allows propagating the spin non-adiabatically, i.e. without accommodations through the contacts. As a consequence this ballistic electron transport requires nanometre size. Notice that we have values over 1000% if there were spin accommodation this value never could be achieved. A tentative scenario for the explanation of this activated behaviour could be based in the existence of set of levels defined by each contact. These are distributed within a wide range of energies in such a way that for each temperature, different subsets of levels contribute to the transport. This is supported because the MR does not depend much with temperature, beside the resistance change by orders of magnitude. In fact the data of Fig.5 can be put in agreement with a bimodal distribution of excited states with peaks around 500K and 1350K.

The work presented here has played with large statistical ensembles of contacts and therefore the MR, the coercive fields, etc are an average. It seems reasonable to think that maybe nancontacts with tremendously high MR values that average their values with other with lower values. The same should happen with other relevant physical properties, as for example coercive fields. These may be handled at wish by changing the sample geometry and number of particles. More work is being done in this direction. Finally we would like to say that the route we have found for MR may not be unique and that probably many more cases exist in nature. Not only that, but the results may depend also of the starting powders. The way we have tried has not been accidental but has been obtained through a set of previous experiment that gave us indications. This is a natural way in which nature provides us with the initial basic units, which is a completely different way respect to the artificially fabricated thin films, perhaps some materials produced following our procedure, cannot be grown in this way, just because nature does not like it. We believe the method presented here is versatile, at hand and easy to perform. Maybe it is the time to start thinking in the direction of nanocontacts produced between nanoparticles, in which oxides and surface properties are provided and tuned by nature. The set of data presented here realize a solid and ample spectrum of physical magnitudes and have been tested over many samples and experimental conditions that should be important in magnetoelectronic applications. It seems that values to the level of 1000% in MR, in a reproducible and stable way, with a very low noise to signal ratio, of the order of 0.1%, will bring the expectations to an impressive level for the future of magnetic devices. More work using other geometries and materials is under way.

Authors acknowledge J.M. De Teresa, L. Morellon and P.A. Algarabel for valuable discussions and a critical reading of the manuscript. Also M. Muñoz, M. L. Sartorelli and A. A. Pasa are acknowledge for measuring and checking the samples that were sent to them. This work has been partially supported by the EU-FP6-STRP-NMP BMR Project. M.R Ibarra received support from the CICyT-MAT2002.

**Figure captions**

**Fig.1.-a)** Scheme of one model of multiple nanocontacts device (MUND). A column of Fe fine particles was compressed on a metallic plate electrode by a tin pin electrode and then glued. **1b)** Resistance variation as function of the injected current. Large variation of R effect is observed at 0 field. This can only be an electronic effect since the field is zero. Notice that at 5000 Oe the R is almost constant with injected current and therefore the MR varies from 1000 to 500% approximately as the current changes from 1 to 10 µA. The continuous lines are to visualize. Inset: typical room temperature MR loops. The first half loop of the virgin sample have a large variation of resistance and after that becomes stabilized.

**Fig 2.-** Dependence of the room temperature magnetoresistance as a function of the overall oxygen content on the particles. The higher MR values correspond to 220 hours of milling in vacuum. The lower values in oxygen correspond to powders that were reduced and then milled while the higher oxygen content correspond to powder milled under an oxygen atmosphere. Values of 300-500% MR and 2-3% (0.25%) oxygen content, where obtained for powders milled in an N atmosphere.

**Fig. 3.-** I/V results at two selected temperatures (273 K and 103 K) at the maximum applied field of 12 Tesla and at the coercive field. Inset: Magnetoresitance of a high field MR cycle (-12,12 Tesla). For the sake of clarity only the range -3,+3Tesla is shown.

**Fig. 4.-** Superposition of 400 room temperature magnetoresistance loops measured over four months period. The resistance can change but the MR remains constant. Inset.- Room temperature time stability of the signal/noise rate with and without applied field.

**Fig 5.-** Thermal cycling under a field of 20 KOe. The line is a fitting of the data using a bimodal distribution of excited states centered at 500 and 1350K.

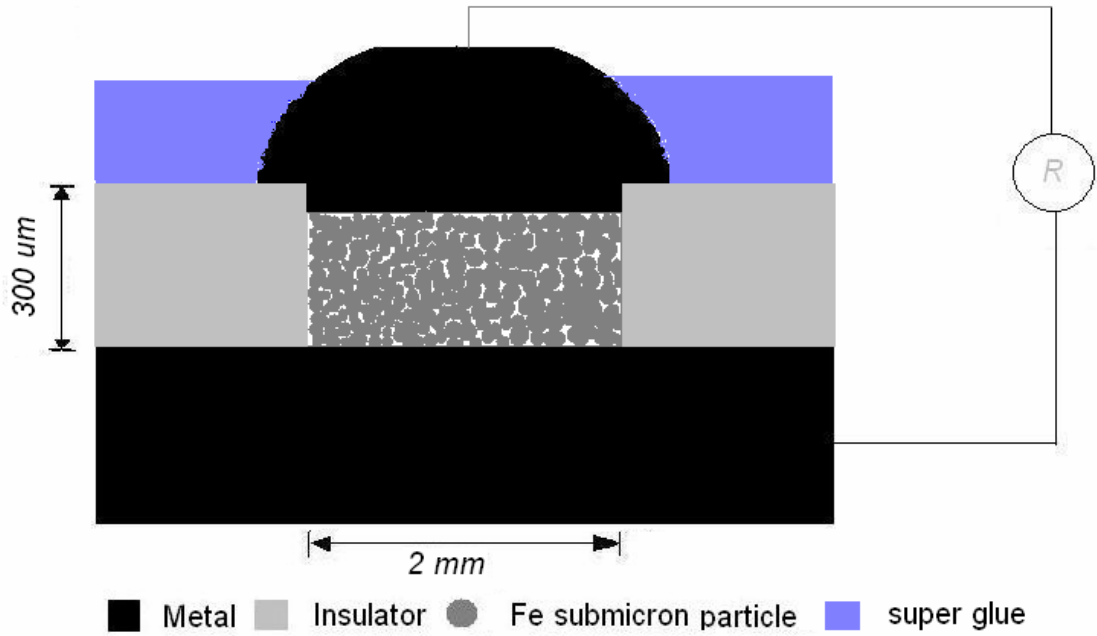

Figure 1 A

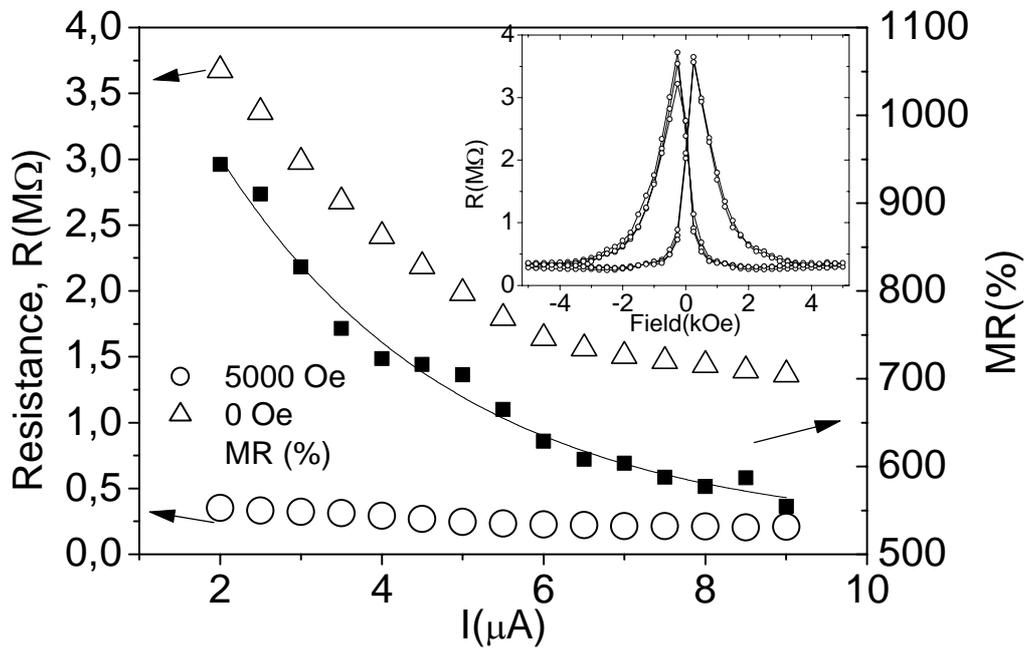

Figure 1 B

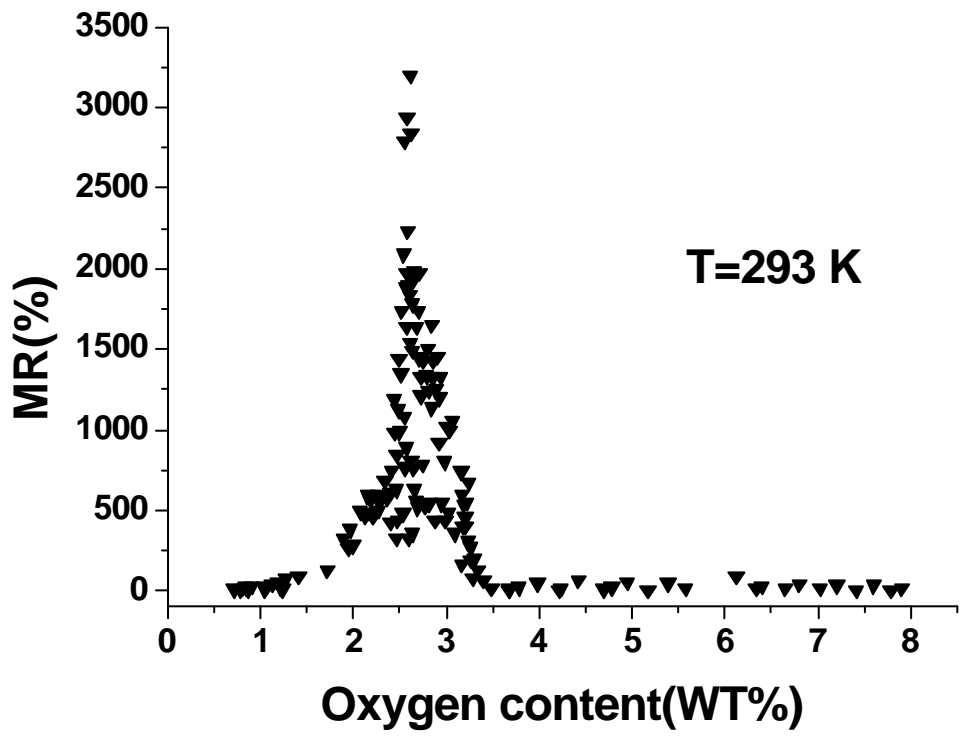

Figure 2

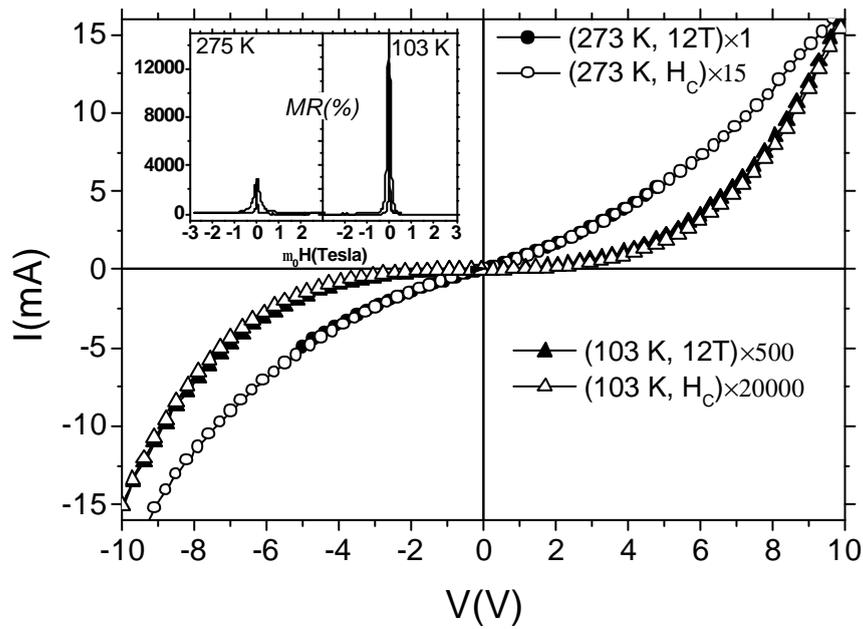

Figure 3

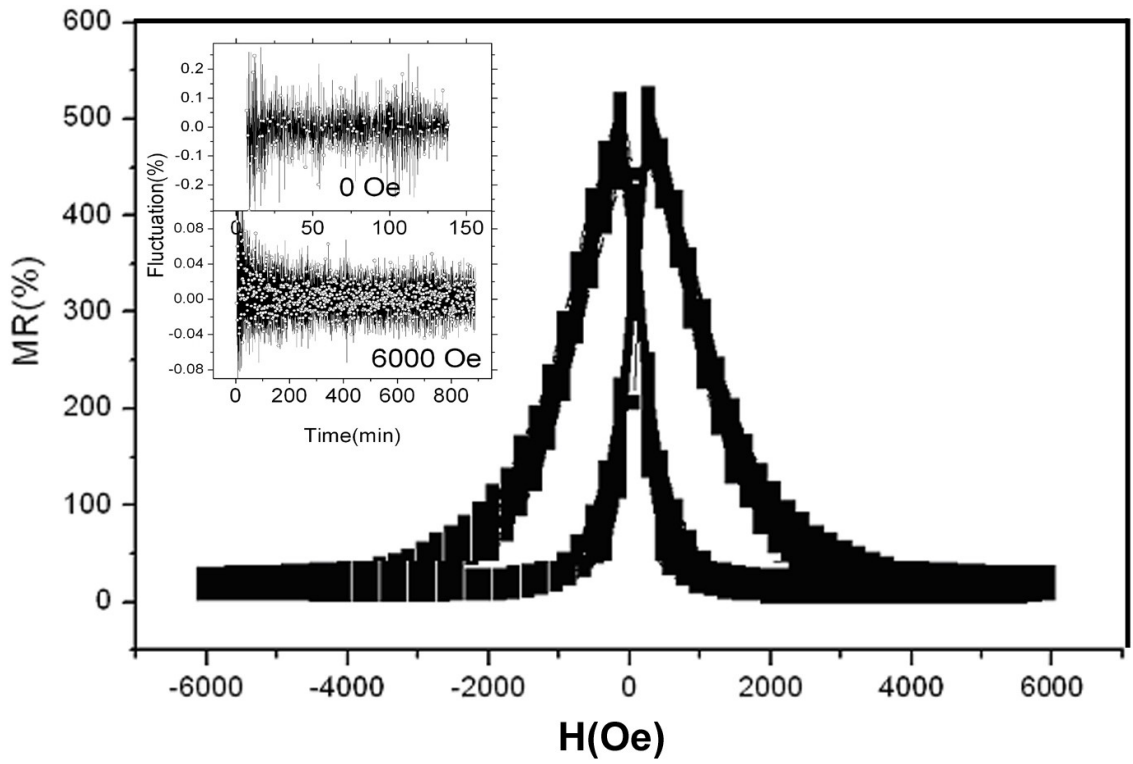

Figure 4

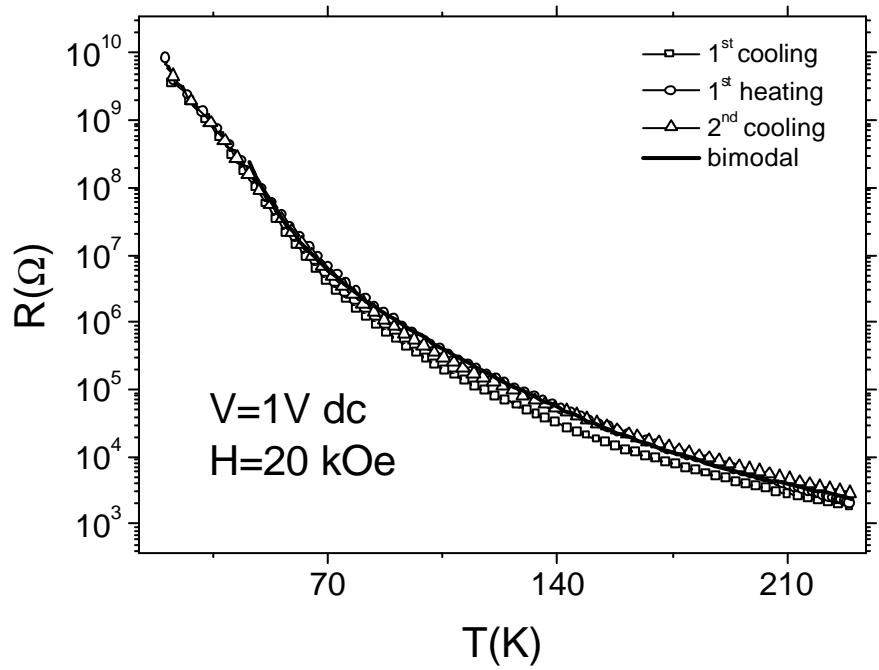

Figure 5